# SubwayMeshDTN: Exploring Opportunistic Delay Tolerant Routing Protocols when Disseminating Emergency Alerts on a Smart City Subway Network

Bruce Estivil
School of Computer Science
The University of Nottingham
psybe1@nottingham.ac.uk

Milena Radenkovic
School of Computer Science
The University of Nottingham
milena.radenkovic@nottingham.ac.uk

*Abstract*—This paper seeks to understand the effectiveness of using multi-dimensional opportunistic delay-tolerant network (DTN) routing protocols, specifically Epidemic and MaxProp, in the context of New York City's (NYC) metropolitan subway network. We examine how efficiently emergency messages spread through mobile, self-configuring, edge-based movement patterns on the train network to understand and propose solutions for improving communication in subterranean environments. Since DTNs are able to store, carry and forward messages through intermediate edges, this paper benchmarks both Wi-Fi and Bluetooth topologies to compare and critically evaluate movement patterns, latency, overheads and delivery rates on pseudo-realistic underground traces. We also show that the accordion effect is predominant in these networks, and therefore, the most effective protocol configurations vary.

## I. Introduction (Heading 1)

Multiple complex problems originate from the unreliability of traditional cellular networks during emergencies due to network overcrowding, complex architecture and varying depths of subway tunnels that hinder the effectiveness and penetration of incumbent communication methods. The rapid development of the Internet of Things (IoT) also necessitates edge computing, which is more scalable, flexible, and enhances security. With artificial intelligence developments, adaptive and decentralised systems are being implemented at breakneck speed to improve the reliability of networks.

Traditional networking systems are negatively impacted during disasters due to hardware damage and the limited bandwidth capacity of adjacent systems to handle the sudden increase in data transmission.

The promising technique of using heterogeneous mobile edges to store, carry and forward messages based on Delay-Tolerant Networks (DTNs) is invaluable for urban transit systems because it accounts for protocol differences and frequent disconnections based on edge mobility and sparseness.

This paper conducts multi-dimensional evaluations of DTN vehicle ad-hoc networking (VANET) scenarios by utilising mobile devices (e.g. trains) to account for the Accordion effect[1]. This occurs frequently on a smart city subway system like the Metropolitan Transportation Authority's (MTA) New York Subway, where there is a dichotomy of connections in the network; either brief or prolonged links are established between edges during rapid changes in network topology. The sharp contrast encapsulates the challenge of ensuring meaningful data exchange occurs during momentary connectivity windows since topology expands and shrinks over time, affecting connection times and data transfer opportunities between edges. In this case, efficient routing protocols are critical, as they must quickly discern and exploit these short-lived opportunities to advance data using intermediate hops, even in the constantly shifting landscape of edge encounters. This paper explores how the Epidemic and MaxProp protocols handle these complexities.

## II. Related Work

Existing research investigates this problem area, scrutinising Mobile Ad-hoc Networks (MANETs) [2] against benchmarks [3] and understanding how they handle malicious edges during disasters [4]. Emerging frameworks have been proposed to reduce energy costs whilst keeping high dissemination rates during emergencies [5]. Mobile Opportunistic Disconnection Tolerant Networks and Systems (MODiToNeS) [19] has been proposed which allows developing distributed architectures that support real-time multi-layer and multi-dimensional communication, so that DTN can be better integrated with mobile social and transportation systems and enable real time communications and services in heterogeneous networks between drones and vehicles [6].

Epidemic [7] is a routing protocol that uses the flooding concept. Data is passed across all encountered edges to propagate messages through the network. In practice, replicating messages causes implementation issues such as significant bandwidth usage. To avoid this, termination conditions are usually present, such as a time-to-live (TTL) or a hop limit, after which the message is assumed to have expired. These methods prevent undeliverable or invalid messages from consuming significant bandwidth and buffer space due to redundant message copies.

MaxProp [8] is another routing protocol that uses the flooding concept but is more sophisticated by enhancing message delivery probabilities, managing buffer space more efficiently, and decreasing delivery times. MaxProp updates internal node data about network topology and node mobility to understand network environments better. Acknowledged messages are generated when a message reaches its destination(s) and is propagated back through the network, where successful reception deletes the message from the sender edge.

Overall, Epidemic retains simplicity in design and implementation, with very high delivery rates and no need for network topology information for the protocol to operate efficiently. The downsides are high bandwidth consumption, increasing network congestion, and substantial buffer space requirements to store all copies of the data.

MaxProp better prioritises messages with adaptive routing based on network topology and better buffer management. However, it's more complex to implement, based on the extended resources required, which can lead to scalability issues in large urban environments. MaxProp's capability makes it better suited for handling and disseminating alerts, as critical messages would be delayed under the Epidemic protocol, which treats all messages equally.

The ONE simulator [9] is a tool in the study of delay-tolerant networks that enables accurate and scalable performance analysis with real-world traces, that is highly suitable for researching network connectivity in difficult environments such as rural, disaster and dense urban areas. ONE simulator models edge movement, routing, and message handling to emulate the store-and-forward paradigm. The simulator produces delivery probability, latency, overheads and hop count data for each pseudo-realistic model.

## III. SUBWAY MESHDTN TRIALS

The New York Subway is one of the largest subway systems in the world, with over 400 stations [10]. This study examines Epidemic and MaxProp with different network interfaces, ranges, and bandwidths to compare their effectiveness against these benchmarks.

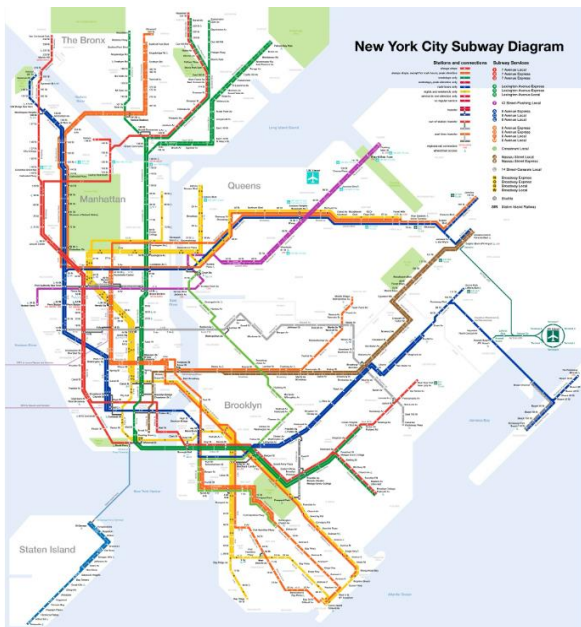

Fig. 1.  Metropolitan Transportation Authority (MTA) subway map [11]

We fuse multiple heterogeneous datasets with a large number of stations and interchanges on the system, especially in Lower Manhattan (Figure 1), to create complex, pseudo-realistic, static edges and allow rich and detailed models of the New York Subway system that enable communication patterns and account for dynamic changes in passenger volumes. With an average of 3.2 million daily rides across the network [12], there are naturally varied connectivity levels in the subway system, with highly dense areas like Lower Manhattan having strong connectivity, whilst outer boroughs Brooklyn and Queens experience intermittent cellular connectivity. This inconsistency poses a challenge for real-time communication, which is crucial for managing passenger flow and disseminating service alerts.

### A. Enabling emergency communication in New York's MTA system

Delay Tolerant Networks (DTN) within the subway system have many use cases unique to New York's subway challenges.

Since all the lines pass through Midtown Manhattan, interchanging between train services in this area causes bottlenecks, especially during peak hours. Even with frequent express services, only stopping at key stations, designed to alleviate high passenger numbers, faults and delays can significantly disrupt schedules. Therefore DTN usage is crucial in managing congestion and enhancing the passenger experience. Real-time updates about delays or service disruptions can be communicated efficiently using DTNs and even reach edges in areas with poor network coverage. This system ensures passengers are informed about delays and provided with alternative routes. For example, during delays at key stations, messages are relayed to passengers before they reach affected stations, providing rerouting suggestions. This proactive approach maintains good passenger flow throughout the network, improving efficiency 24/7. Utilising DTN networks on the subway relates to emergency communication handling. Critical dependence on cellular networking can lead to information gaps, and therefore, DTNs would become an instrumental bridge. Emergency alerts, instructions, or updates can be transmitted by passenger devices, acting as DTN edges, as they move through the system on trains and stations, ensuring continuous information flow even when direct, cellular-based networking is unavailable. This approach is vital where immediate communication is required, i.e. evacuations or service suspensions. The system remains fully connected and delivers a responsive and passenger-centric subway service by ensuring passengers in all areas receive timely, relevant, and accurate information.

With higher subway ridership at off-peak hours (16% increase on peak hours) [12], most train interchanges occur during these periods. Diversions, maintenance and alternative routing requirements occur at an increased rate on nights and weekends, so ensuring customers always understand their options is crucial to efficient journeys.

### B. Development of SubwayMeshDTN

To further investigate, the 'Subway Lines' dataset by NYC Open Data [13] was obtained to modelled and build continuous edge movement throughout the map.

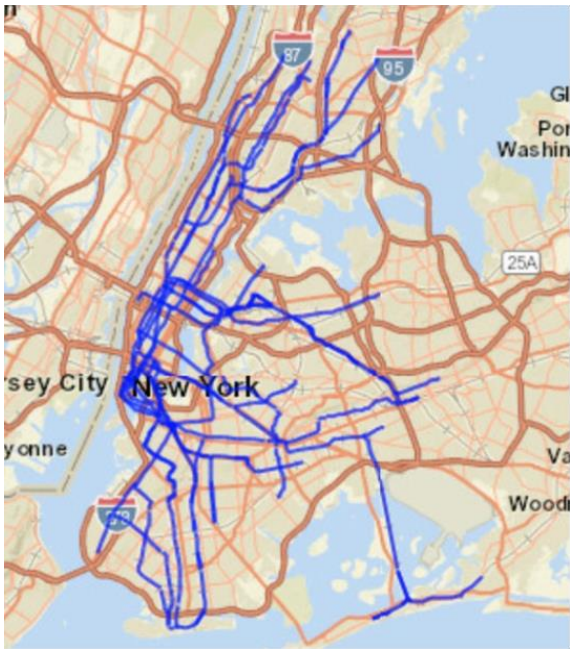

Fig. 2. New York City's Open Data -'Subway Lines'

Multiple heterogeneous datasets are fused to display an interconnected map representing the subway system, allowing bidirectional edge movement. There are two groups of edges, with local trains having the prefix 'L', express trains having the prefix 'E', and any event actions having the 'e' prefix.

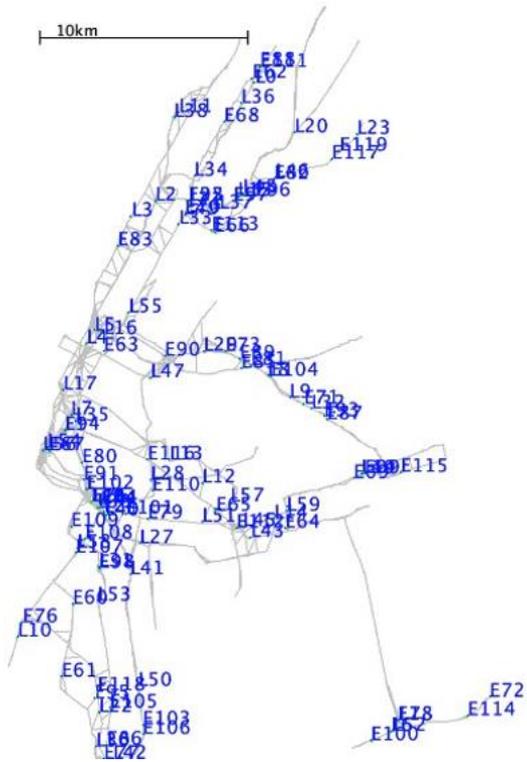

Fig. 3. MTA Open Data - ONE Simulator

Epidemic and MaxProp routing protocol tests are run multiple times for different network interfaces, ranges and bandwidths. For tests using Bluetooth, the edge range is set at 10 meters and for tests with Wi-Fi range, it's set at 30 meters. All tests contain 60 edges per group (120 total), where the max speed local edges (trains) are set to 17.4mph, the average train speed on the network [14], and express edges are set to 55mph, the network's top speed.

## IV. MULTI-CRITERIA, MULTI-PROTOCOL CRITICAL ANALYSIS OF SUBWAYMESHDTN

### A. Evaluation of SubwayMeshDTN

In the face of dynamic topology and communication patterns, the experiment yielded the following results. Message delivery is viewed as a service or emergency alert completing enough hops to propagate a substantial part of the network.

TABLE I. EPIDEMIC (BLUETOOTH) EDGES

| | |
|---|---|
| Alerts (Delivered / Created) | 104 / 521 |
| Alert Delivery Rate | 20% |
| Alert Delivery Lanency (average) | 13291 seconds |
| Node Hop (Initiated / Completed) | 15765 / 6719 |
| Hop Completion Rate | 42.6% |

TABLE II. MAXPROP (BLUETOOTH) EDGES

| | |
|---|---|
| Alerts (Delivered / Created) | 99 / 521 |
| Alert Delivery Rate | 19% |
| Alert Delivery Lanency (average) | 12529 seconds |
| Node Hop (Initiated / Completed) | 14309 / 5117 |
| Hop Completion Rate | 35.8% |

The Bluetooth results indicate that the Epidemic protocol outperformed MaxProp in alert delivery and hop completion rates. The constrained range of 10 meters significantly hindered MaxProp's ability to perceive the network topology, thus impeding the capacity to develop synchronously efficient routing patterns. However, MaxProp still demonstrated superior end-to-end latency metrics from message dissemination to recorded delivery.

TABLE III. EPIDEMIC (WIFI) EDGES

| | |
|---|---|
| Alerts (Delivered / Created) | 173 / 521 |
| Alert Delivery Rate | 33.2% |
| Alert Delivery Lanency (average) | 10783 seconds |
| Node Hop (Initiated / Completed) | 85428 / 82990 |
| Hop Completion Rate | 97.1% |

TABLE IV. MAXPROP (WIFI) EDGES

| | |
|---|---|
| Alerts (Delivered / Created) | 281 / 521 |
| Alert Delivery Rate | 53.93% |
| Alert Delivery Lanency (average) | 9359 seconds |
| Node Hop (Initiated / Completed) | 77528 / 74987 |
| Hop Completion Rate | 96.7% |

With Wi-Fi edges, both protocols exhibited significantly improved hop completion rates, mitigating the impacts of the accordion effect. Since the Epidemic protocol indiscriminately floods the network, it facilitated a higher frequency of data transfer opportunities, increasing the quantity of initiated and completed hops. Conversely, MaxProp had approximately 15% better average latency, rendering it more effective for efficient data dissemination.

Both Wi-Fi iterations of the protocols performed better than their counterparts, which is evident from the higher alert delivery rates and lower latency times observed with Wi-Fi edges. This is the effect of extended range and transfer speeds available, allowing these edges to connect to other edges over longer distances. In density-populated subway networks, especially experiencing the accordion effect, these capabilities are particularly advantageous in ensuring efficient data transfer since edges have more time to complete each "hop".

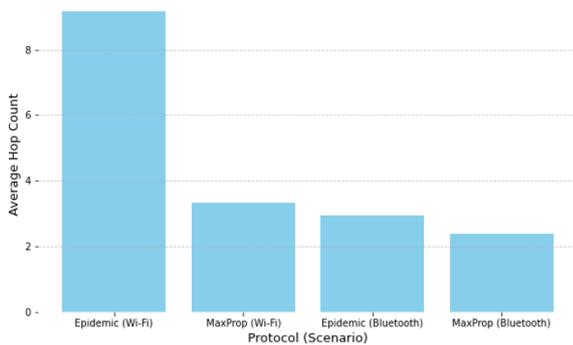

Fig. 4. Average hop count for each scenario

A performance comparison of Epidemic and MaxProp routing protocols inside the New York City underground system revealed considerable differences, particularly in the Wi-Fi-based scenarios. The Epidemic (Wi-Fi) scenario exhibited a significantly higher hop rate, over double that of the next highest. However, the message delivery rate fell around 20% short of the MaxProp (Wi-Fi) scenario's result.

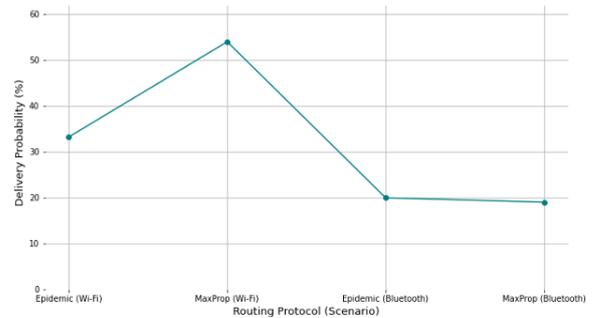

Fig. 5. Delivery probability for each scenario

This difference in delivery probability is due to the two protocols' data handling algorithms. The Epidemic protocol uses a flooding concept to disseminate information to every accessible edge. This strategy ensures wide information dissemination but does not prioritise or discriminate in its distribution, which leads to inefficiencies and redundancies in message delivery. The MaxProp protocol, on the other hand, takes a more deliberate strategy, passing information to edges when there is a higher likelihood of successful delivery than the present edge. This strategy saves bandwidth while increasing message delivery rates for appropriate recipients. This conclusion is also observed in the Bluetooth scenario, with Epidemic having a 1% better delivery rate than its counterpart.

Latency is an important metric when determining the best protocols to use. In emergency or service disruptions on the subway system, timely information delivery to affected parties allows more informed decisions and is crucial to providing efficient service.

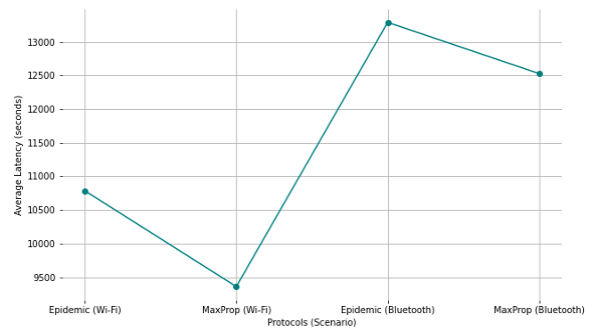

Fig. 6. Latency comparison for each scenario

Figure 6 shows the MaxProp scenarios had lower latencies for both edge types but a more significant gap between MaxProp (Wi-Fi) and Epidemic (Wi-Fi). This result also relates to MaxProp's probabilistic routing mechanism and Wi-Fi's technological advantages, which result in a more streamlined and faster message delivery process, achieving lower latency.

The overhead ratio is another key metric to understand how efficient the network is, as measured by how little extra traffic is generated for each message sent.

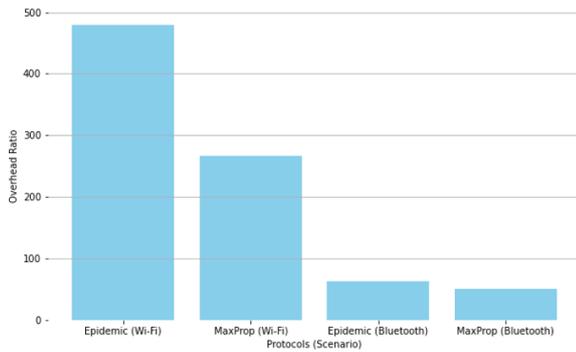

Fig. 7. Overhead ration for each scenario

Figure 7 shows that Epidemic (Wi-Fi) has nearly double the next highest overhead ratio, and both Wi-Fi scenarios contain high overhead due to extensive message propagation and, therefore, increased overhead ratio. Both Epidemic scenarios also contain higher overheads, as when messages flood indiscriminately, more copies of each message are created and transmitted. As with latency and delivery rates, MaxProp's ability to prioritise messages and flood when necessary makes it suitable for highly dense and crowded environments because alerts are delivered more efficiently, using less network capacity.

## V. Discussion and Further Research

For the NYC subway, MaxProp (Wi-Fi) results proved best suited for this environment. It allows for targeted transmission of alerts to customers most likely to be impacted by specific situations. For example, by setting a target delivery edge further along a subway line, MaxProp routing ensures that passengers in that direction receive relevant messages. This targeted approach is particularly beneficial in managing real-time situations like service disruptions or emergency alerts, where disseminating information to the most affected passengers is crucial.

Furthermore, the MaxProp protocol's efficiency in message delivery aligns well with the system's operational needs. By reducing the number of redundant messages and focusing on high-probability delivery paths, MaxProp enhances the overall effectiveness of the communication system.

While Epidemic demonstrates a robust capacity for message propagation through its high hop rate, MaxProp, particularly in the Wi-Fi scenario, offers a more efficient and targeted approach to message delivery in the NYC subway system.

Since traditional networks can be ineffective due to intermittent connectivity, installing edges strategically across the network allows the system to communicate alerts with customers effectively. On the other hand, extreme scenarios such as severely delayed or diverted trains reduce the effectiveness and therefore, the system must be maintained consistently to operate at its full capability.

In future research, we aim to tackle potential security and privacy vulnerabilities inherent in DTNs. Since these protocols pass information to available edges, there exists an inherent risk of data interception and unauthorised access. Furthermore, the energy consumption to maintain these networks is extremely high. Other research has explored and provided valuable insights into optimising energy consumption, for example, within Fog Networks [15]. However, detailed security, privacy, and energy consumption considerations are outside the scope of this paper.

Overall, these networks can be a vital mechanism during natural disasters. There have been numerous examples of destroyed traditional communication methods hindering connectivity and amplifying the effects of disasters. In 2011, the Japan Tsunami resulted in large-scale infrastructure damage, including communication networks [16]. The broken system hindered emergency response efforts and the local government's spread of important information. Similarly, the Boston Marathon bombing in 2013 caused 281 injuries [17], but also severely impacted cellular networks due to sudden surges in calls and messages. This congestion restricted effective communication for first responders and was a catalyst for the spread of misinformation, which caused additional panic and confusion [18].

For both situations, these networks would've been a resilient communication method capable of operating despite infrastructure damage. In Japan, information about evacuation routes, relief centres and other safety messages could've been communicated efficiently. In Boston, they would've established a clear communication channel with first responders to prevent panic and misinformation. Critical information could've been shared between responding teams regarding further possible attacks and casualties. Post-event, during the inquiry phases, the infrastructure would've recorded and stored 'black-box' style information to help uncover the motives and create input measures to prevent such tragedies from reoccurring.

Further research into enhancing energy efficiency, constructing resilient infrastructure, and applying blockchain technology to fortify data integrity collectively contributes to improved quality of life. These advancements effectively address the needs of citizens, aligning with broader objectives of urban development and sustainability within smart cities.